\documentclass[12pt]{article}
\pdfoutput=1
\usepackage{color}
\usepackage{graphicx}
\usepackage{amsmath}
\usepackage{slashed}

\input xy
\xyoption{all}
\xyoption{web}

\newlength{\abstractwidth}

\renewcommand{\thefootnote}{\fnsymbol{footnote}}
\renewcommand{\thanks}[1]{\footnote{#1}}
\newcommand{\starttext}{
\setcounter{footnote}{0}
\renewcommand{\thefootnote}{\arabic{footnote}}}

\newcommand{\bea}{\begin{eqnarray}}
\newcommand{\eea}{\end{eqnarray}}
\newcommand{\ee}{\end{equation}}
\newcommand{\be}{\begin{equation}}
\newcommand{\ea}{\end{array}}

\newcommand{\half}{\frac{1}{2}}

\begin{document}
\starttext
\setcounter{footnote}{0}

\bigskip

\begin{center}

{\Large \bf A note on interface solutions in higher-spin gravity}
\vskip 0.6in

{ \bf Michael Gutperle}

\vskip .2in

{ \sl Department of Physics and Astronomy }\\
{\sl University of California, Los Angeles, CA 90095, USA}\\
{\tt  gutperle@physics.edu; }

\vskip .7in

{\bf Abstract}

\end{center}

\medskip

In this paper we construct holographic interface solutions in the Chern-Simons formulation of higher-spin gravity. It is shown that the global regularity of the gauge connection imposes conditions on the conserved currents consistent with interface being totally transmissive or topological.
An interface solution connecting the $W_3$ and $W_3^{(2)}$ vacuum is constructed in the  $SL(3,R)$ higher spin gravity. An  interface solution  corresponding to turning on relevant deformations in higher spin $SL(4,R)$ gravity   is  presented.

\newpage


\baselineskip=17pt
\setcounter{equation}{0}
\setcounter{footnote}{0}

\newpage

\section{Introduction}
\setcounter{equation}{0}
\label{sec1}

Recently,  there has been a renewed interest higher spin gravity primarily in the formulation due to Vasiliev and collaborators (see  e.g. \cite{Vasiliev:2000rn} for a review).  This was partially  motivated by the  conjecture  of Klebanov and Polyakov  \cite{Klebanov:2002ja}  that such theories in four dimensions are dual to the large $N$ limit of  three dimensional $O(N)$
vector models.

In three dimensions higher spin theories can be implemented as  a Chern-Simons gauge theory, 
generalizing the formulation of  three dimensional gravity  with a negative  cosmological constant as $SL(2,R)
\times SL(2,R)$ Chern-Simons gauge theory \cite{Witten:1988hc,Achucarro:1987vz}.  The simplest case is obtained by replacing the $SL(2,R)
$ gauge group by $SL(N,R)$ and describes a theory of  higher spin fields with  spins up to  
$s=2,3,\cdots,N$. On the dual CFT side, Drinfeld-Sokolov reduction \cite{Drinfeld:1984qv} gives a realization of $W_N$ symmetry. This
generalizes the Virasoro  symmetry of the pure gravity case \cite{Campoleoni:2010zq,Henneaux:2010xg,Campoleoni:2011hg}.  The three dimensional version of Vasiliev 
theory is constructed as a  Chern-Simons  based on the infinite dimensional gauge algebra 
$hs(\lambda)$ \cite{Blencowe:1988gj,Bergshoeff:1989ns,Pope:1989sr}. The $SL(N,R)$ gauge group  leads to a purely topological theory in the bulk  with no propagating physical degrees of freedom. On the other hand for  the infinite dimensional gauge algebra $hs(\lambda)$, it is possible to consistently couple propagating matter to the higher spin gravity theory.  The bulk theory was conjectured  \cite{Gaberdiel:2010pz,Gaberdiel:2011wb,Gaberdiel:2012ku}  to be dual to a t'Hooft like limit of coset  $W_N$ minimal models
\be
{SU(N)_k \otimes SU(N)_1  \over SU(N)_{k+1}}  \quad k,N\to \infty, \quad \lambda ={N\over N+k} \;\; {\rm fixed}
\ee
Many interesting solutions of higher spin gravity have been found and interpreted on the CFT side. Examples are generalizations of the BTZ black hole carrying higher spin charge and conical defects.  A novel feature of higher spin 
gravity manifests itself in these examples: in general the metric is not invariant under higher spin gauge transformations and purely geometric characterizations of what constitutes a black hole \cite{Gutperle:2011kf,Ammon:2011nk,Castro:2011fm,Ammon:2012wc} or a 
conical singularity  \cite{Castro:2011iw} are not gauge invariant. A new criterion for regularity involving holonomy conditions on the Chern-Simons connections was proposed in \cite{Gutperle:2011kf}.

The  asymptotic symmetry of the AdS vacuum in $SL(N,R)\times SL(N,R)$ higher spin gravity  depends on the embedding of the $SL(2,R)$ sub-algebra inside $SL(N,R)$. For the principal 
embedding one  obtains $W_N$ symmetry, whereas for non-principal embeddings other higher 
spin algebras such as the Bershadsky-Polyakov $W_3^{(2)}$ algebra \cite{Polyakov:1989dm,Bershadsky:1990bg} occur.  See \cite{Castro:2012bc,Afshar:2012hc} for recent work on such theories. 

It is also possible to consider solutions of higher spin gravity where the asymptotic geometry is not $AdS_3$, examples in the literature contain  Lifshitz, warped $AdS$ and Lobachevsky geometries  \cite{Gary:2012ms,Afshar:2012nk}.  One common feature of such solutions is that they   only work  for non-principal 
embeddings of $SL(2,R)$, in  which  additional singlet sectors are present in the branching of   $SL(2,R)$ representations.

The goal of this note is to construct new solutions of higher spin gravity which realize interface \footnote{See \cite{Karch:2000gx} for an early construction of interface and boundary CFTs in the context of AdS/CFT.} or Janus geometries.  A Janus solution \cite{Bak:2003jk}
 uses an $AdS_{d}$ slicing of $AdS_{d+1}$ to construct a gravity solution which is dual to a defect or interface CFT. The simplest example can be constructed in type IIB, utilizing an $AdS_4$ slicing of $AdS_5$ where  only the dilaton and the metric have a nontrivial profile. This solution is dual to $3+1$ dimensional  $N=4$ $SYM$ 
 theory with  the Yang-Mills coupling  jumping across a $2+1$ dimensional interface. In recent years 
other solutions   have been constructed in supergravity including supersymmetric Janus solutions 
 in type IIB  supergravity  \cite{Clark:2005te,D'Hoker:2007xy,D'Hoker:2007xz}, M-theory \cite{D'Hoker:2008wc} and six dimensional supergravity  \cite{Chiodaroli:2009yw,Chiodaroli:2011nr}. A feature of all of 
 these solutions is that the interface is one between two CFTs at different points in their moduli 
 space.   More generally interfaces can connect different CFTs. A special class of interface CFTs are so called topological interfaces   \cite{Petkova:2000ip,Bachas:2001vj}, which are totally transmissive \cite{Quella:2006de}. 

The structure  and the main results of the paper are  as follows: In section \ref{sec2} we introduce our 
notation and conventions and present the Chern-Simons gauge connections which produce  
$AdS_3$ in  $AdS_2$ slicing coordinates. In the simplest setting of pure gravity we 
show that the natural boundary conditions on the stress tensor are characteristic of a totally transmissive or topological interface. In section \ref{sec3} we consider the simplest higher spin theory, 
namely $SL(3,R)\times SL(3,R)$  Chern-Simons theory and construct a solution which realizes an interface between the two different $AdS_3$ vacua corresponding to the principal and non-principal embeddings, namely the $W_3$ and $W_3^{(2)}$ vacuum. In 
section \ref{sec4} we consider the $SL(4,R)\times SL(4,R)$ theory  and construct a Janus-like interface solution between the same CFT on both sides. The presence of $SL(2,R)$  singlets in the 
embedding, i.e. the fact that we have to utilize a non-principal embedding, is essential in this construction. We close the paper in section \ref{sec5} with a discussion  of our results and direction for  further work. Some technical details are relegated to appendices.

\section{Interface solution in Chern-Simons gravity}
\setcounter{equation}{0}
\label{sec2}
The Chern-Simons formulation of three dimensional (higher-spin) gravity is based on two copies of Chern-Simons action at level $k$ and $-k$. 
\be\label{chernsimonsa}
S=S_{CS}[A]-S_{CS}[\bar A]
\ee 
where
\be
S_{CS}[A]= {k\over 4\pi}  \int {\rm tr}\Big( A\wedge dA+{2\over 3} A\wedge A\wedge A\Big)
\ee
The equations of motion are simply flatness conditions,
\be
F=dA+ A\wedge A=0, \quad \quad \bar F=d\bar A+ \bar A\wedge \bar A=0
\ee
 In this section we discuss the Chern-Simons formulation of pure gravity which uses $SL(2,R)\times SL(2,R)$ gauge 
connections.  Since the construction of higher spin solution always involves an embedding of 
$SL(2,R)$ which realizes the spin 2 sector, the results obtained here are the starting point for the higher spin generalizations.
The vielbein and spin connection are related to the CS gauge fields as follows
\be\label{vielbein}
e={1\over 2} (A-\bar A), \quad \omega=\half (A+\bar A)
\ee 
and the metric is obtained from the gauge connections by
\be\label{gmetdef}
g_{\mu\nu} =2 {\rm tr} (e_\mu e_\nu)
\ee
The starting point of constructing interface solutions is to express the $AdS_3$ metric as an $AdS_2$ slicing
\be\label{janslice}
ds_2^2= d\mu^2 + \cosh^2\mu \left( {  - dt^2+dz^2 \over z^2}\right) 
\ee
Here $\mu\in[-\infty,\infty]$ is the slicing coordinate and the boundary of $AdS_3$ is made up from two half spaces at $\mu\to \pm \infty$ glued together at a real line at $z\to 0$.

The $SL(2,R)$ gauge connection which realizes the metric (\ref{janslice}) is  most easily constructed using light cone coordinates, defining  $x^+={1\over 2} (z+t), \quad x^-={1\over 2}(z-t)$.
\be\label{bigadef}
A= b^{-1} a b + b^{-1}db, \quad \bar A = b \bar a b^{-1} + b \;d  (b^{-1}) 
\ee
with $b=e^{\mu l_0}$and the $\mu$ independent connections $a_\pm  dx^\pm$ and $ \bar a_\pm  dx^\pm$ are given by
\bea\label{adsback}
a_+ &=& {1\over z}\big( l_+ +l_0\big), \quad  a_- = {1\over z} \big(l_--l_0\big) \nonumber \\
\bar a_+ &=& {1\over z} \big(- l_++ l_0\big), \quad 
\bar a_-= {1\over z}\big(-l_- -l_0\big)
\eea
It is easy to check that using (\ref{bigadef}) and  (\ref{gmetdef}) the connection (\ref{adsback}) indeed reproduces the metric (\ref{janslice}).

\subsection{Asymptotic symmetries}\label{sltwoasym}

The the metric formulation of $AdS_3$ gravity, the conformal transformations given by the Virasoro 
algebra are constructed from diffeomorphisms which preserve the asymptotic $AdS$ metric. In the  Chern-Simons formulation the asymptotic symmetries are  realized by gauge transformations
\be\label{gauget}
\delta A = d\Lambda + [A, \Lambda], \quad \delta \bar A = d\bar \Lambda + [A, \bar \Lambda]
\ee
preserving the asymptotic structure of the gauge connection. In the following we will consider a special class of gauge transformations adapted to the $\mu$ slicing in  (\ref{bigadef})
\be
\Lambda = b^{-1}\lambda(x^+,x^-) b, \quad \bar \Lambda =b \bar \lambda(x^+,x^-) b^{-1}
\ee
Here we will display the argument only for the gauge connection $A$ and only summarize the results for the gauge connection $\bar A$ at the end of this section.
We consider a deformation of the $AdS$ background  given by a  ``right" connection parameterized by 
\bea\label{rightcon}
A^R_+& = & b^{-1}{1\over z}\big( l_+ +l_0+ z^2 {\cal L}^R(x^+)l_-\big)b\; = \; {1\over z}\big( e^\mu  l_+ +l_0+ z^2  e^{-\mu} {\cal L}^R(x^+)l_-\big)\nonumber \\
A^R_- &=& b^{-1} {1\over z} \big(l_--l_0\big) b\; = \; {1\over z} \big(e^{-\mu}l_--l_0\big) 
\eea
It is easy to verify that this connection satisfies the flatness condition for an  ${\cal L}^R$ which depends only on $x^+$. Furthermore in the limit $\mu\to \infty$, the ${\cal L}^R$-dependent term in $A^R_+$ decays  exponentially and does not alter the asymptotic $AdS$ geometry. The gauge transformation which does not alter the asymptotic geometry in the $\mu\to \infty$ limit is generated by an infinitesimal gauge transformation parameter $\epsilon^R(x^+)$
\be
\lambda^R=  {\epsilon^R \over z} l_+ + \Big( {2 \epsilon^R \over z} -  \partial_+ \epsilon^R\Big) l_0 + \Big( -{ \epsilon^R\over z} +  \partial_+ \epsilon^R -{z\over2} \partial_+^2\epsilon^R  + {2\pi z {\cal L}^R \over k} \epsilon^R\Big)l_-
\ee
Under this gauge transformation the connection (\ref{rightcon}) remains the same apart from ${\cal L}^R$ which transforms as
\be
\delta {\cal L}^R = \epsilon^R \partial_+ {\cal L}^R+ 2  \partial_+ \epsilon^R {\cal L}^R-{k \over 4\pi} \partial_+^3  \epsilon^R
\ee
Which is exactly the infinitesimal version of the Virasoro algebra if we identify ${\cal L}^R$ with the stress tensor. Recall that the $AdS_3$ boundary contains three components: two half spaces reached by taking $\mu\to \pm \infty$ and a real line at which the two half spaces are glued together at $z\to 0$.  Hence the above analysis is only valid for the ``right" half space reached by taking $\mu\to \infty$. In addition, the ${\cal L}^R$ dependent term  in (\ref{rightcon}) is dominant when one takes $\mu\to -\infty$.

To remedy this situation we introduce a second ``left" connection dependent on ${\cal L}^L(x^-)$
\bea
A^L_+& =&  {1\over z} b^{-1} \big(  l_+ +l_0\big)b \;= \;{1\over z}  \big( e^{-\mu}  l_+ +l_0\big) , \nonumber \\
A^L_- &=& {1\over z} b^{-1}\big(l_--l_0+ z^2  {\cal L}^L(x^-)l_+\big) b\; =\; {1\over z} \big(e^{-\mu} l_--l_0+ z^2  e^{\mu} {\cal L}^L(x^-)l_+\big)
 \eea
In this case the  ${\cal L}^L$ decays as $\mu\to -\infty$ as one approaches the ``left" boundary component.
The gauge transformation which leaves the asymptotic AdS geometry unchanged is parameterized by a function $\epsilon^L(x^-)$.
\be
\lambda^L=  {\epsilon^L \over z} l_- - \Big( {2 \epsilon^L\over z} -  \partial_-\epsilon^L\Big) l_0 + \Big( -{ \epsilon^L\over z} +  \partial_-\epsilon^L -{z\over2} \partial_-^2 \epsilon^L  + {2\pi z {\cal L}^L \over k} \epsilon^L\Big)l_+
\ee
and under this gauge transformation ${\cal L}^L$ transforms as
\be
\delta {\cal L}^L = \epsilon^L \partial_- {\cal L}^L+ 2  \partial_-\epsilon^L {\cal L}^L-{k \over 4\pi} \partial_-^3 \epsilon^L 
\ee
Hence we have constructed a left moving and right moving copy of the Virasoro algebra via Drinfeld-Sokolov reduction. However the left and the right moving symmetries are only well defined on the left and right boundary component respectively. In the next section we will argue that the stress tensors have to satisfy matching conditions for the gauge connections to be globally defined.
\subsection{Interface matching conditions}

In order to obtain a gauge connection which is well defined for all $\mu$ and asymptotes to the $AdS$  values for both $\mu \to +\infty$ and $\mu \to -\infty$  the $R$ and $L$ gauge connections have to be related by a gauge transformation at some finite value of $\mu=\mu_0$.  It is possible to perform this gauge transformation on the $a_\pm$ fields. This implies that  the  $\mu$, where the left and right connections are matched, can be chosen arbitrarily. Hence, we are looking for a gauge transformation $g(x^+,x^-)$  such that
\bea\label{gaugerl}
a^R_{\pm} &=&  g^{-1} \; a^L_\pm \; g + g^{-1} \partial_\pm  g  
\eea
We limit ourselves to considering an  infinitesimal  stress energy tensor deformation ${\cal L}^{L,R} = o(\epsilon)$ 
\be
{\cal L}^R(x^+) = \epsilon\;  T^R(x^+)+o(\epsilon^2) ,\quad    {\cal L}^L(x^-) =  \epsilon \;  \bar T^L(x^-)+o(\epsilon^2)
\ee
and gauge transformations 
\bea
g&=& 1+ \epsilon \Big( H _+ (x^+,x^-) l_+ + H_-(x^+,x^-)  l_- + H_1(x^+,x^-) l_0\Big)+ o(\epsilon^2) \nonumber \\
\bar g &=& 1+ \epsilon \Big( G _+ (x^+,x^-) l_+ +G_-(x^+,x^-)  l_- + G_1(x^+,x^-) l_0\Big)+ o(\epsilon^2) \nonumber 
\eea
The matching condition (\ref{gaugerl}) can be satisfied  with the following choices
\bea
H_+ &=& {1\over 2} \left( {h(x^+) + \bar h(x^-) \over z} - \bar h'(x^-)+ {z\over 2}    \bar h''(x^-) \right) \nonumber \\
H_- &=& {1\over 2} \left( -{h(x^+) +  \bar h(x^-) \over z} + h'(x^+)- {z\over 2}    h''(x^+) \right) \nonumber \\
H_1&=& \left( {h(x^+) + \bar h(x^-) \over z}- {1\over 2} h'(x^+) - {1\over 2} \bar h'(x^-)\right)
\eea
where the parameters $h,\bar h$ satisfy
\be\label{htrel}
\partial_+^3 h  =-{2\pi\over k} T^R(x^+) , \quad  \partial_-^3 \bar h = - {2\pi\over k} \bar T^L(x^-)
\ee
Note that a matching gauge transformation $g$  has to be finite, and due to the $1/z$  factors this condition imposes a boundary condition on $h, \bar h$ 
\be
\Big( h(x^+) + \bar h(x^-)\Big) |_{z=0}    =0
\ee
Using the relation (\ref{htrel}) and the fact that at $z=0$ the light cone derivatives satisfy  $\partial_+ = -\partial_-$, we find that    the boundary condition on $T$
\be
\Big(T^R(x^+)- \bar T^L(x^-)\Big)\big|_{z=0}=0
\ee
An analogous calculation for the gauge transformations of the barred gauge connection $\bar A$ leads to the condition 
\be
\Big(\bar T^R(x^-)- T^L(x^+)\Big)\big|_{z=0}=0
\ee
Hence the matching conditions of the fluctuations of the stress energy tensor are the ones of a completely transmissive or topological interface.

\section{RG-Interface solution in $SL(3,R)$  higher spin gravity}
\setcounter{equation}{0}
\label{sec3}
The Chern-Simons formulation of three dimensional AdS gravity can be generalized  by replacing $SL(2,R)$ by other gauge groups. The simplest case is given by $SL(3,R)$. This theory was studied in detail in \cite{Campoleoni:2010zq} and it was shown that the CS theory is equivalent to $AdS$ gravity coupled to a massless spin three field. Using the  expression of the vielbein (\ref{vielbein})  in terms of the connection,   the metric and spin 3 field can be expressed as
\be\label{metform}
g_{\mu\nu}={1\over 2} tr(e_\mu e_\nu), \quad  \quad \phi_{\mu\nu \rho} ={1\over 6} tr(e_{(\mu} e_\nu e_{\rho)})
\ee

The $SL(3,R)$ connections which reproduce  the $AdS_2$ slicing of $AdS_3$ are of the same form as in (\ref{adsback}) with the  matrix  form of the generators   given in appendix \ref{conven}. The symmetry of the boundary theory can be determined by repeating the  Drinfeld-Sokolov  reduction  described in section \ref{sltwoasym} with the connection
\bea\label{slthreer}
a^R_+ &=&  {1\over z} \Big( l_1+ l_0 -{2\pi\over k} z^2 {\cal L}^R(x^+)l_{-}-{\pi \over 2k} z^3 {\cal W}^R(x^+) \; w_{-2}\Big), \quad \quad 
a_-^R =   {1\over z} \Big( l_{-1}- l_0 \Big)
\eea
The gauge transformations (\ref{gauget}) which leave the asymptotic form of the AdS geometry unchanged in the limit $\mu\to +\infty$ generate $W_3$ transformations where ${\cal L}$ transforms as the stress tensor and ${\cal W}$ as the spin three $W$ generator (see e.g. \cite{Gutperle:2011kf} for details).

As discussed in the previous section in the connection $A^R$ obtained from $a^R$, the terms containing  ${\cal L}$  and  ${\cal W}$ are only subleading in the $\mu\to +\infty$ limit, corresponding to one half-space. As before we have to introduce a second gauge connection $a^L$ which covers the other boundary half space, reached by taking $\mu\to -\infty$.

\bea\label{slthreel}
a^L_+ &=& {1\over z}\Big (l_1+l_0\Big), \quad a^L_- = {1\over z}\Big(-l_{-}- l_0 +{2\pi\over k} z^2 {\cal \bar L}^L(x^-)l_{+}+{\pi \over 2k} z^3 {\cal \bar W}^L(x^-) \; w_{2}\Big) 
\eea
As in section \ref{sec2} we demand that the two gauge connections are gauge equivalent in an overlapping region of finite $\mu$. The finiteness of such a gauge transformation imposes conditions relating ${\cal L}^R(x^+), {\cal W}^R(x^+)$ to ${\cal L}^L(x^-), {\cal W}^L(x^-)$ at the interface. Since the ${\cal L},{\cal W}$  are dual   to the stress tensor and spin three current in the dual CFT we can obtain interface conditions relating the left and right moving quantities at $z=0$.  We relegate some  details  of calculation to appendix \ref{slthreegauge} and only give the final result here:
\bea
\Big(T^R(x^+) -\bar T^L(x^-)\Big)\Big|_{z=0} & =&0, \quad \quad \Big(W^R(x^+) -\bar W^L(x^-)\Big)\Big|_{z=0} =0 \nonumber\\
\Big(\bar  T^R(x^-) -T^L(x^+)\Big)\Big|_{z=0} &=&0, \quad \quad \Big(\bar W^R(x^-) -W^L(x^+)\Big)\Big|_{z=0} =0,
\eea
The boundary condition for the stress tensor imply that the interface is topological and the boundary conditions on the conserved spin three current are  the natural generalization  for $W_3$ CFTs. The study of topological interfaces in $W_N$ CFTs would be very interesting but to our knowledge this has not yet been done.   

\subsection{RG-flow interface}
For $SL(3,R)$ there are two inequivalent embeddings of $SL(2,R)$. The principal embedding corresponds to choosing $l_{\pm 1}, l_0$ as the $SL(2,R)$ generators.  The principal embedding  produces an asymptotic AdS geometry which has a CFT with $W_3$ symmetry. On the other hand a non-principal embedding corresponds to choosing (rescaled) $w_{\pm 2}, l_0$  as the $SL(2,R)$ generators.
Using these generators to construct the asymptotic AdS geometry produces a CFT with  the Polyakov-Bershadsky $W_3^{(2)}$ symmetry algebra. The black hole solutions of \cite{Gutperle:2011kf} were interpreted \cite{Ammon:2011nk} (in a particular gauge)
 as  wormhole solutions corresponding to a RG flow between the two two CFTs. 

Here we present a simple modification of the $AdS_2$ slicing connection which produces an interface between the $W_3$ vacuum  on one half space and the $W_3^{(2)}$ vacuum on the other.

\bea
A_+ &=& {1\over z} \Big(e^\mu l_1+ l_0\Big), \quad   A_- ={1\over z} \Big( c \;e^{-2\mu} w_{-2} -e^{-\mu} l_{-1}-l_0\Big), \quad A_\mu=l_0 \nonumber \\
\bar A_+ &=& {1\over z} \Big(-c\; e^{-2\mu}  w_2 + l_0-l_1 \Big), \quad   \bar A_- ={1\over z} \Big( -e^{\mu} l_{-1}-l_0\Big), \quad \bar  A_\mu=-l_0 \nonumber \\
\eea

\begin{figure}[htbp]
\begin{center}
\includegraphics[scale=0.50]{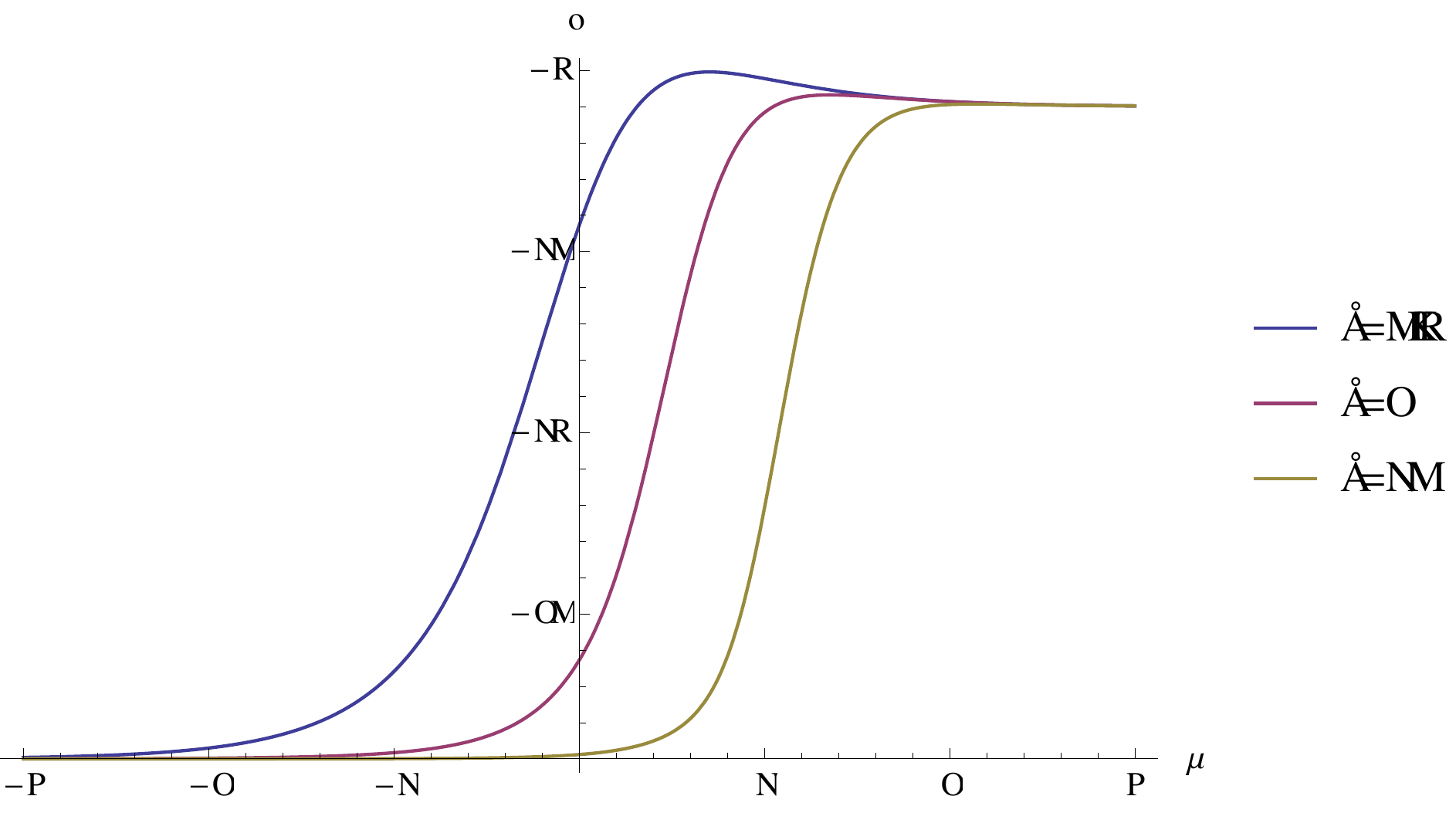}
\caption{Plot of Ricci scalar (\ref{riccisc}) for various values of $c$.}
\label{figurea}
\end{center}
\end{figure}

Note that the deformation depends on a single parameter $c$. The leading terms of the connection in the $\mu\to \infty $  limit depend on $l_{\pm1}$ and hence produce  the $W_3$ vacuum and the leading term in the $\mu\to -\infty$ limit depend on $w_{\pm2}$ and produce the $W_3^{(2)}$ vacuum. Using (\ref{metform}) the metric takes the following form
\bea\label{intmetb}
ds^2 = d\mu^2 -\big( e^{-4\mu} c^2 + \cosh^2\mu  \big) {1\over z^2}  \Big(-dt^2 + dz^2\Big) 
\eea

The non-vanishing components of the spin 3 field are given by
\be
\phi_{ttz}=-c{e^{2\mu} \cosh^2\mu\over z^3}, \quad \quad \phi_{z,z,z}=3c{e^{2\mu} \cosh^2\mu\over z^3}
\ee
The Ricci scalar calculated from the metric (\ref{intmetb}) is given by
\be\label{riccisc}
R= {e^{8 \mu} \cosh^2\mu\big(1+ 3 \cosh2\mu\big) -4c^2\big( 6 c^2+2 e^{2\mu}+4 e^{4\mu}+ 3 e^{6\mu}\big)\over (c^2 + e^{4\mu}\cosh^2 \mu)^2}
\ee

One concludes from (\ref{riccisc}) and figure \ref{figurea} that the solution interpolates between two asymptotic $AdS$ regions with cosmological constant $\Lambda = -8$ for $\mu\to -\infty$ and $\Lambda=-2$ for $\mu\to +\infty$  corresponding to the $W^3$ invariant and $W_3^{(2)}$ invariant  vacua. It is tempting to interpret  this solution as a realization of the idea of a RG-flow interface originally proposed in \cite{Gaiotto:2012np}.

\section{Interface solution in $SL(4,R)$ higher spin gravity}
\setcounter{equation}{0}
\label{sec4}
A Janus solution in gravity is an interface solution between CFTs at different points in their moduli space. In particular this means that the central charge $c$ is the same on both sides of the interface. Such a solution is different from the RG-flow interface presented in section \ref{sec3}.  As for the solutions of CS higher spin gravity which are non-AdS, like Lifshitz and warped AdS \cite{Gary:2012ms,Afshar:2012nk}, we were unable to find Janus solutions for embedding which do not contain singlets.
The simplest non-rotating solution we were able to construct utilized the $(2,2)$ embedding of $SL(4,R)$. The  $SL(2,R)$ generators are denoted by $l_{-1},l_0,l_{+1}$. The branching and the explicit form of the generators are displayed in appendix \ref{conven}. The $(2,2)$ embedding contains three $SL(2,R)$ triplets   and three singlets, which will be denoted as  $s_{-1},s_0,s_{+1}$. For all embeddings the spin $s$ of the $SL(2,R)$ representation is related to the spin  $\Sigma$ of the associated  field in $AdS_3$ by $\Sigma=s+1$.  The conformal dimension of the conserved  CFT current dual to the spin $S$ field is $\Delta=s+1$.

As an explicit example we consider the following one parameter family of connections $A,\bar A$ which are of the form  (\ref{bigadef}) with $b=e^{\mu  l_0}$
  \bea\label{conjanusa}
A_+&=&{1\over z} \Big(-e^\mu l_1 +l_0+ s_0 +(1-2e^{x_1})s_1\Big) \nonumber\\
A_-&=& {1\over z} \Big(e^{-\mu} l_{-1} -l_0-(1-e^{x_1})s_0 +{e^{x_1}\over 2} s_{-1} -(1-{1\over 2} e^{x_1})s_1\Big)\nonumber\\
\bar A_+&=& {1\over z} \Big(e^{-\mu} l_1 +l_0+ s_0 +(1-2e^{-x_1})s_1\Big)\nonumber\\
\bar A_-&=& {1\over z} \Big(-e^{\mu} l_{-1} -l_0-(1-e^{-x_1})s_0 +{e^{-x_1}\over 2} s_{-1} -(1-{1\over 2} e^{-x_1})s_1\Big)
\eea
It is straightforward to verify that the connection (\ref{conjanusa}) satisfies the equation of motion. 
The metric (\ref{gmetdef}) obtained from the connection is given by
\be
ds^2 = {1\over 2} \Big\{d\mu^2 +\big(\cosh^2 \mu  - \sinh^2 x_1  \big){-dt^2+dz^2\over z^2} \Big\}
\ee
Note that the metric is a deformation of the $AdS_3$ vacuum as can be seen from the plot of the Ricci-scalar as a function of $x_1$

\begin{figure}[htbp]
\begin{center}
\includegraphics[scale=0.60]{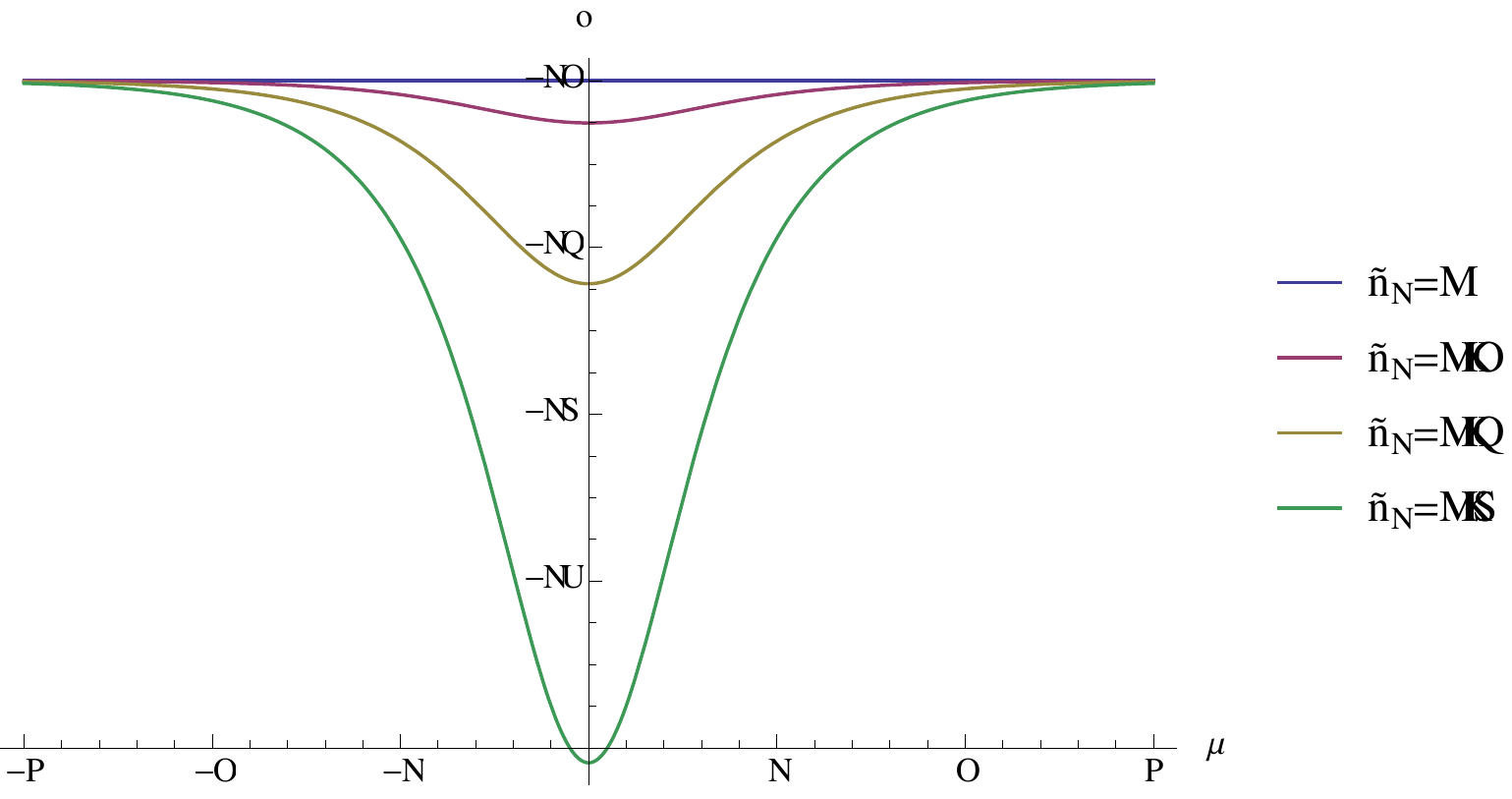}
\caption{Plot of Ricci-scalar (\ref{ricciscb}) for various values of $x_1$.}
\label{figurea}
\end{center}
\end{figure}

\be\label{ricciscb}
R={8\cosh (2\mu) \big(2\cosh 2x_1 -5\big) + 8 \cosh (2x_1) -6 \cosh (4\mu) -26\over (2 +\cosh \big(2\mu) -2\cosh (2x_1)\big)^2}
\ee

The geometry in both in the $\mu\to \pm \infty$ limit asymptotes to $AdS$ space with the same cosmological constant corresponding to a vacuum for the W algebra related to the $(2,2)$ embedding. Note that the metric develops a naked singularity for the critical value $x_1={1\over 2 }\cosh^{-1} 3$.  As discussed above, the singlets are dual to spin one currents. Since the conformal dimension of these currents is one, turning  on these fields corresponds to a IR relevant deformation which grows away from the boundary.  Such solutions were found in \cite{Gutperle:2012hy} for ordinary $AdS$ gravity coupled to a massive scalar and display qualitatively similar features.

\section{Discussion}
\setcounter{equation}{0}
\label{sec5}

In the present paper we have constructed new solutions of higher spin gravity in the Chern-Simons formulation, which are dual to  interface CFTs.  These solutions are constructed by deforming the Chern-Simons connections of an $AdS_3$ in $AdS_2$ slicing coordinates.  Due to the fact that 
there are two boundary components the asymptotic Virasoro (and $W_N$) symmetries are realized separately  on the two boundary components and have to be glued together by an Chern-Simons gauge 
transformation.  This existence of such a gauge transformation imposes conditions on the stress tensor and  other conserved  currents along the interface which connects the two boundary 
components.   The conditions imposed on the stress tensor are totally transmissive and correspond to topological interface solutions. We also constructed solutions which are interface theories with CFTs deformed by IR relevant spatially dependent couplings first found in \cite{Gutperle:2012hy}. 
In the example we have found it was necessary to employ $SL(2,R)$ singlet fields, which only appear in non-principal embeddings. 

 Furthermore we have constructed Janus like solutions corresponding to interfaces between the $W_3$ and $W_3^{(2)}$ vacua the $SL(3,R)$ higher spin theory. Such solutions can be interpreted as interfaces of theories related by RG flows in the spirit of \cite{Gaiotto:2012np}. In this context it would be very interesting to consider generalizations of these construction to the case of $hs(\lambda)$ higher spin theory since a concrete dual  exists  \cite{Gaberdiel:2010pz,Gaberdiel:2011wb,Gaberdiel:2012ku}  as a coset CFT and the methods of classifying and constructing topological interfaces  developed in \cite{Gaiotto:2012np} could be compared to the bulk construction outlined in the present paper. Another interesting and important feature of $hs(\lambda)$ theory lies in the fact that it is possible to consistently couple
propagating scalars to the theory. A particularly simple case might be the $\lambda=1$ where one has a massless scalar and one might expect the possibility to construct a dual of a Janus solution \cite{Bak:2003jk}. Furthermore the dual theory is one of free fermions in which interface conditions can be investigated using CFT techniques \cite{Bachas:2001vj,Oshikawa:1996dj}. Another interesting quantity which can be calculated for a interface CFT is the entanglement entropy across the interface \cite{Calabrese:2004eu}. This quantity  can be related to the boundary entropy of the interface \cite{Affleck:1991tk}. This entanglement entropy was calculated holographically in \cite{Azeyanagi:2007qj,Chiodaroli:2010ur,Chiodaroli:2010mv} for Janus solutions in supergravity.  It would be interesting to investigate wether the entanglement entropy can be calculated from the Chern-Simons side. We leave these questions for future work.

\bigskip

\noindent{\Large \bf Acknowledgements}

\bigskip

This  work was in part supported  by NSF grant PHY-07-57702. I am grateful to Joshua Samani for useful conversations and comments on a draft.

\bigskip

\appendix

\section{Conventions}\label{conven}
\setcounter{equation}{0}

In this appendix explicit expressions for the $SL(N,R)$ generators used in the body of the text are presented.

\medskip

\noindent $\bullet$ The $SL(2,R)$ generators  are given by
\be
l_{+}= \left(
\begin{array}{cc}
 0 & 1    \\
  0& 0    
\end{array}
\right),\quad l_-=  \left(
\begin{array}{cc}
 0 & 0     \\
  1& 0
\end{array}
\right),\quad l_0=\half \left(
\begin{array}{cc}
 -1 & 0     \\
  0& 1
\end{array}
\right) 
\ee
and satisfy the following commutation relations
\be\label{lplml1}
[l_+,l_0]=l_+, \quad [l_-,l_0]=-l_-, \quad [l_+,l_-]=-2 l_0
\ee

\medskip

\noindent $\bullet$  The $SL(3,R)$ generators employed in section \ref{sec3} are given by
\bea
l_{+}&=& \left(
\begin{array}{ccc}
 0 & 0 &  0 \\
  1& 0&0\\
  0&1&0    
\end{array}
\right), \quad l_{0}= \left(
\begin{array}{ccc}
 1 & 0 &  0 \\
  0& 0&0\\
  0&0&-1    
\end{array}
\right), \quad l_{-}= \left(
\begin{array}{ccc}
 0 & -2 &  0 \\
  0& 0&-2\\
  0&0&0    
\end{array}
\right)\nonumber \\
\nonumber\\
w_{2}&=& \left(
\begin{array}{ccc}
 0 & 0 &  0 \\
  0& 0&0\\
  2&0&0    
\end{array}
\right), \quad w_{1}= \left(
\begin{array}{ccc}
 0 & 0 &  0 \\
  1& 0&0\\
  0&-1&0    
\end{array}
\right), \quad w_{0}= {2\over 3} \left(
\begin{array}{ccc}
 1 & 0 &  0 \\
  0& -2&0\\
  0&0&1   
\end{array}
\right)\nonumber\\
\nonumber\\
w_{-1}&=& \left(
\begin{array}{ccc}
 0 & -2 &  0 \\
  0& 0&2\\
  0&0&0    
\end{array}
\right), \quad
 w_{-2}= \left(
\begin{array}{ccc}
 0 & 0 &  8 \\
  0& 0&0\\
  0&0&0    
\end{array}
\right)
\eea
There are two inequivalent embeddings of $SL(2,R)$ in $SL(3,R)$. The principal embedding chooses $l_{+}, l_0,l_-$. The $w_i,i=-2,\cdots,2$ form a spin 2 representation of $SL(2,R)$. 
A second $SL(2,R)$ embedding is given by  ${\small {1\over 4}} W_2, {\small {1\over 2}} L_0, -{\small {1\over 4}} W_{-2}$ where the other generators split into a singlet generated by $W_0$ and two spin half multiplets generated by $W_1,L_{-1}$ and $L_1, W_{-1}$ respectively.

\medskip

\noindent $\bullet$ The basis of $SL(4,R)$ generators used in section \ref{sec4} is taken from \cite{Gary:2012ms} (a slightly different but equivalent basis was used in \cite{Tan:2011tj}). Note that the inequivalent $SL(2,R)$ embeddings  in $SL(N,R)$ are labelled by a partition $(n_1,n_2,\cdots ,n_k)$ of $N$. For the construction of the Janus solution in section \ref{sec4} we need the $n_1=2,n_2=2$ embedding. The generators split into three spin one  triplets and three singlets. The $SL(2,R)$ generators are given by

\bea
l_0=
{1\over 2} \left(
\begin{array}{cccc}
1  & 0  &0   &0\\
 0 &  1 &0   &0\\
 0 &0   &-1   &0\\
 0 & 0 &0 &-1
\end{array}
\right), \;l_+=
\left(
\begin{array}{cccc}
 0 &0   &0   &0\\
 0 & 0  & 0  &0\\
  1& 0  &0   &0\\
 0 &1&0&0
\end{array}
\right),\; l_-=
\left(
\begin{array}{cccc}
  0&0   &-1   &0\\
 0 & 0  &0   &-1\\
  0& 0  & 0  &0\\
 0 &0&0&0
\end{array}
\right)
\eea
The singlets are given by
\bea
s^{[0]}=
{1\over 2} \left(
\begin{array}{cccc}
1  & 0  &0   &0\\
 0 &  -1 &0   &0\\
 0 &0   &1   &0\\
 0 & 0 & 0&-1
\end{array}
\right),\; s^{[+]}=
\left(
\begin{array}{cccc}
0  &0   & 0  &0\\
 1 & 0  & 0  &0\\
0  & 0  & 0  &0\\
 0 &0&1&-
\end{array}
\right),\;s^{[-]}=
\left(
\begin{array}{cccc}
  0& -1  &0   &0\\
  0& 0  & 0  &0\\
0  & 0  &0   &-1\\
  0&0&0&0
\end{array}
\right),
\eea

In addition there are three triplet generators which are not needed in the main body of the paper, there explicit form can be found in \cite{Gary:2012ms}.

\section{Gauge transformation for $SL(3,R)$ interface}\label{slthreegauge}
\setcounter{equation}{0}

The right and left gauge connections (\ref{slthreer}) and (\ref{slthreel}) are given by
\bea\label{slthreer}
a^R_+ &=&  {1\over z} \Big( l_1+ l_0 -{2\pi\over k} z^2 {\cal L}^R(x^+)l_{-}-{\pi \over 2k} z^3 {\cal W}^R(x^+) \; w_{-2}\Big), \quad \quad 
a_-^R =   {1\over z} \Big( l_{-1}- l_0 \Big)
\eea
and 
\bea\label{slthreel}
a^L_+ &=& {1\over z}\Big (l_1+l_0\Big), \quad a^L_- = {1\over z}\Big(-l_{-}- l_0 +{2\pi\over k} z^2 {\cal \bar L}^L(x^-)l_{+}+{\pi \over 2k} z^3 {\cal \bar W}^L(x^-) \; w_{2}\Big) 
\eea
We demand that the left and right gauge connections are gauge equivalent in an overlapping region of finite $\mu$, i.e. 
\bea\label{slthreegaugea}
a^R_+ &=& g^{-1}  a^L_+ g- g^{-1}\partial_+ g, \quad \quad a^R_- = g^{-1}  a^L_- g- g^{-1}\partial_- g
\eea
Here we only display the unbarred gauge field, the barred gauge fields obey an analogous relation.
For a finite $SL(3,R)$ group element which can depend on $x^+,x^-$ but is independent of $\mu$. Repeating the argument presented in section \ref{sec2} we limit ourselves to infinitesimal deformations 
\be
{\cal L}^{L,R}  = \epsilon\; T^{L,R}, \quad\quad {\cal W}^{L,R} =\epsilon\; W^{L,R}
\ee
where we identify $T$ with the stress tensor and $W_3$ with the generator. The infinitesimal gauge transformation is given by 
\bea
g&=&1+ \epsilon\Big\{ H_1 l_{+} + H_2 l_0+H_3 l_{-1}+ H_4 w_{2} +H_5 w_{1}+H_6 w_0+H_7 w_{-1}+H_8 w_{-2}\Big\}+o(\epsilon^2)\nonumber\\
\eea
The matching condition (\ref{slthreegaugea}) can be solved in terms of two right moving  functions $f_+(x^+), g_+(x^+)$ and two left moving functions $f_-(x^-),g_-(x^-)$.
\bea
H_1&=& {1\over 2z} \big(f_++f_- \big) -{1\over 2}\partial_- f_ -+{1\over 4}  z \partial_-^2 f_-\nonumber \\
H_2&=&{1\over z}\big( f_++f_-\big) -{1\over 2} \big( \partial_+ f_+ +\partial_- f_-\big)\nonumber\\
H_3&=& {1\over 2z} \big(f_++f_- \big) -{1\over 2}\partial_+ f_ ++{1\over 4}  z \partial_+^2 f_+\nonumber\\
H_4&=& {1\over 6 z^2} \big( g_++g_-\big)- {1\over 6z} \partial_- g_-+ {1\over 12} \partial_-^2 g_- -{z\over 36}  \partial_-^3 g_-+ {z^2\over 144} \partial_-^4 g_- \nonumber\\
H_5 &=& {2\over 3 z^2}\big( g_+ + g_-\big)-{1\over 6 z} \big(\partial_+ g_++3 \partial_- g_-\big)+{1\over 6}  \partial_-^2 g_- -{z\over 36} \partial_-^3 g_-  \nonumber\\
H_6&=& {1\over z^2}\big( g_++g_-\big) -{1\over 2 z}\big(\partial_+ g_++\partial_- g_-\big)+ {1\over 12} \big( \partial_+^2 g_++ \partial_-^2 g_-\big)\nonumber \\
H_7 &=& {2\over 3 z^2}\big(g_+ +g_-\big) -{1\over 6 z} \big( \partial_+ g_+ +3 \partial_- g_-\big) +{1\over 6} \partial_+^2 g_+ -{z\over 36} \partial_+^3 g_+ \nonumber\\
H_8 &=&{1\over 6 z^2}\big(g_+ +g_-\big) -{1\over 6 z}\partial_+ g_++{1\over 12} \partial_+^2 g_+-{z\over 36} \partial_+^3 g_++{z^2\over 144} \partial_+^4 g_+
\eea
The functions $f_\pm,g_\pm$ are related to the quantities in the connection (\ref{slthreer}) and (\ref{slthreel}) as follows
\bea\label{twrel}
\partial_+^3 f_+&=& {2\pi \over k} T^R(x^+),\quad \partial_+^5 g_+ = - {72 \pi\over k} W^R(x^+)\nonumber \\
\partial_-^3 f_-&=& {2\pi \over k} \bar T^L(x^-),\quad \partial_-^5 g_- = - {72 \pi\over k} \bar W^L(x^-)
\eea
Imposing the finiteness of the gauge transformation $g$ in the limit $z\to 0$ imposes conditions relating $f^+,g^+$ to $f^-,g^-$  at the interface where $z=0$, i.e.  $x^+=x^-$
\be\label{gfrel}
\Big( g_+(x^+) + g_-(x^-) \Big)\Big|_{z=0}=0, \quad  \Big( f_+(x^+) + f_-(x^-) \Big)\Big|_{z=0}=0,
\ee
Using (\ref{twrel}) and the fact that at $z=0$ one has $\partial_+ =-\partial_-$ the condition (\ref{gfrel}) translates into matching conditions of the stress tensor  T and W current on the two boundaries
\be
\Big(T^R(x^+) -\bar T^L(x^-)\Big)\Big|_{z=0} =0, \quad \quad \Big(W^R(x^+) -\bar W^L(x^-)\Big)\Big|_{z=0} =0,
\ee
A completely analogous calculation for the barred gauge fields produces
\be
\Big(\bar \bar T^R(x^-) -T^L(x^+)\Big)\Big|_{z=0} =0, \quad \quad \Big(\bar W^R(x^-) -W^L(x^+)\Big)\Big|_{z=0} =0,
\ee

\newpage



\begin{thebibliography}{99}

{\small

\bibitem{Vasiliev:2000rn}
  M.~A.~Vasiliev,
  ``Higher spin symmetries, star product and relativistic equations in AdS space,''
  hep-th/0002183.
  
\bibitem{Klebanov:2002ja}
  I.~R.~Klebanov and A.~M.~Polyakov,
  ``AdS dual of the critical O(N) vector model,''
  Phys.\ Lett.\ B {\bf 550} (2002) 213
  [hep-th/0210114].

\bibitem{Witten:1988hc}
  E.~Witten,
  ``(2+1)-Dimensional Gravity as an Exactly Soluble System,''
  Nucl.\ Phys.\ B {\bf 311} (1988) 46.

\bibitem{Achucarro:1987vz}
  A.~Achucarro and P.~K.~Townsend,
  ``A Chern-Simons Action for Three-Dimensional anti-De Sitter Supergravity Theories,''
  Phys.\ Lett.\ B {\bf 180} (1986) 89.

\bibitem{Drinfeld:1984qv}
  V.~G.~Drinfeld and V.~V.~Sokolov,
  ``Lie algebras and equations of Korteweg-de Vries type,''
  J.\ Sov.\ Math.\  {\bf 30} (1984) 1975.

\bibitem{Campoleoni:2010zq}
  A.~Campoleoni, S.~Fredenhagen, S.~Pfenninger and S.~Theisen,
  ``Asymptotic symmetries of three-dimensional gravity coupled to higher-spin fields,''
  JHEP {\bf 1011} (2010) 007
  [arXiv:1008.4744 [hep-th]].

\bibitem{Henneaux:2010xg}
  M.~Henneaux and S.~-J.~Rey,
  ``Nonlinear $W_{infinity}$ as Asymptotic Symmetry of Three-Dimensional Higher Spin Anti-de Sitter Gravity,''
  JHEP {\bf 1012} (2010) 007
  [arXiv:1008.4579 [hep-th]].

\bibitem{Campoleoni:2011hg}
  A.~Campoleoni, S.~Fredenhagen and S.~Pfenninger,
  ``Asymptotic W-symmetries in three-dimensional higher-spin gauge theories,''
  JHEP {\bf 1109} (2011) 113
  [arXiv:1107.0290 [hep-th]].

\bibitem{Blencowe:1988gj}
  M.~P.~Blencowe,
  ``A Consistent Interacting Massless Higher Spin Field Theory In D = (2+1),''
  Class.\ Quant.\ Grav.\  {\bf 6} (1989) 443.

\bibitem{Bergshoeff:1989ns}
  E.~Bergshoeff, M.~P.~Blencowe and K.~S.~Stelle,
  ``Area Preserving Diffeomorphisms And Higher Spin Algebra,''
  Commun.\ Math.\ Phys.\  {\bf 128} (1990) 213.

\bibitem{Pope:1989sr}
  C.~N.~Pope, L.~J.~Romans and X.~Shen,
  ``W(infinity) And The Racah-wigner Algebra,''
  Nucl.\ Phys.\ B {\bf 339} (1990) 191.



\bibitem{Gaberdiel:2010pz}
  M.~R.~Gaberdiel and R.~Gopakumar,
  ``An $AdS_3$ Dual for Minimal Model CFTs,''
  Phys.\ Rev.\ D {\bf 83} (2011) 066007
  [arXiv:1011.2986 [hep-th]].


\bibitem{Gaberdiel:2011wb}
  M.~R.~Gaberdiel and T.~Hartman,
  ``Symmetries of Holographic Minimal Models,''
  JHEP {\bf 1105} (2011) 031
  [arXiv:1101.2910 [hep-th]].

\bibitem{Gaberdiel:2012ku}
  M.~R.~Gaberdiel and R.~Gopakumar,
  ``Triality in Minimal Model Holography,''
  JHEP {\bf 1207} (2012) 127
  [arXiv:1205.2472 [hep-th]].



\bibitem{Gutperle:2011kf}
  M.~Gutperle and P.~Kraus,
  ``Higher Spin Black Holes,''
  JHEP {\bf 1105} (2011) 022
  [arXiv:1103.4304 [hep-th]].
  
\bibitem{Ammon:2011nk}
  M.~Ammon, M.~Gutperle, P.~Kraus and E.~Perlmutter,
  ``Spacetime Geometry in Higher Spin Gravity,''
  JHEP {\bf 1110} (2011) 053
  [arXiv:1106.4788 [hep-th]].
  
\bibitem{Castro:2011fm}
  A.~Castro, E.~Hijano, A.~Lepage-Jutier and A.~Maloney,
  ``Black Holes and Singularity Resolution in Higher Spin Gravity,''
  JHEP {\bf 1201} (2012) 031
  [arXiv:1110.4117 [hep-th]].
  
\bibitem{Ammon:2012wc}
  M.~Ammon, M.~Gutperle, P.~Kraus and E.~Perlmutter,
  ``Black holes in three dimensional higher spin gravity: A review,''
  arXiv:1208.5182 [hep-th].
 
 
   
\bibitem{Castro:2011iw}
  A.~Castro, R.~Gopakumar, M.~Gutperle and J.~Raeymaekers,
  ``Conical Defects in Higher Spin Theories,''
  JHEP {\bf 1202} (2012) 096
  [arXiv:1111.3381 [hep-th]].

 
\bibitem{Polyakov:1989dm}
  A.~M.~Polyakov,
  ``Gauge Transformations and Diffeomorphisms,''
  Int.\ J.\ Mod.\ Phys.\ A {\bf 5} (1990) 833.

\bibitem{Bershadsky:1990bg}
  M.~Bershadsky,
  ``Conformal field theories via Hamiltonian reduction,''
  Commun.\ Math.\ Phys.\  {\bf 139} (1991) 71.




\bibitem{Castro:2012bc}
  A.~Castro, E.~Hijano and A.~Lepage-Jutier,
  ``Unitarity Bounds in $AdS_3$ Higher Spin Gravity,''
  JHEP {\bf 1206} (2012) 001
  [arXiv:1202.4467 [hep-th]].

\bibitem{Afshar:2012hc}
  H.~Afshar, M.~Gary, D.~Grumiller, R.~Rashkov and M.~Riegler,
  ``Semi-classical unitarity in 3-dimensional higher-spin gravity for non-principal embeddings,''
  arXiv:1211.4454 [hep-th].
 
 

 
\bibitem{Gary:2012ms}
  M.~Gary, D.~Grumiller and R.~Rashkov,
  ``Towards non-AdS holography in 3-dimensional higher spin gravity,''
  JHEP {\bf 1203} (2012) 022
  [arXiv:1201.0013 [hep-th]].
 
\bibitem{Afshar:2012nk}
  H.~Afshar, M.~Gary, D.~Grumiller, R.~Rashkov and M.~Riegler,
  ``Non-AdS holography in 3-dimensional higher spin gravity - General recipe and example,''
  JHEP {\bf 1211} (2012) 099
  [arXiv:1209.2860 [hep-th]].
 
 
\bibitem{Kraus:2011ds}
  P.~Kraus and E.~Perlmutter,
  ``Partition functions of higher spin black holes and their CFT duals,''
  JHEP {\bf 1111} (2011) 061
  [arXiv:1108.2567 [hep-th]].
  
\bibitem{Karch:2000gx}
  A.~Karch and L.~Randall,
  ``Open and closed string interpretation of SUSY CFT's on branes with boundaries,''
  JHEP {\bf 0106} (2001) 063
  [hep-th/0105132].
  
\bibitem{Bak:2003jk}
  D.~Bak, M.~Gutperle and S.~Hirano,
  ``A Dilatonic deformation of AdS(5) and its field theory dual,''
  JHEP {\bf 0305} (2003) 072
  [hep-th/0304129].
  
\bibitem{Clark:2005te}
  A.~Clark and A.~Karch,
  ``Super Janus,''
  JHEP {\bf 0510} (2005) 094
  [hep-th/0506265].


  
\bibitem{D'Hoker:2007xy}
  E.~D'Hoker, J.~Estes and M.~Gutperle,
  ``Exact half-BPS Type IIB interface solutions. I. Local solution and supersymmetric Janus,''
  JHEP {\bf 0706} (2007) 021
  [arXiv:0705.0022 [hep-th]].
  
\bibitem{D'Hoker:2007xz}
  E.~D'Hoker, J.~Estes and M.~Gutperle,
  ``Exact half-BPS Type IIB interface solutions. II. Flux solutions and multi-Janus,''
  JHEP {\bf 0706} (2007) 022
  [arXiv:0705.0024 [hep-th]].
  
\bibitem{D'Hoker:2008wc}
  E.~D'Hoker, J.~Estes, M.~Gutperle and D.~Krym,
  ``Exact Half-BPS Flux Solutions in M-theory. I: Local Solutions,''
  JHEP {\bf 0808} (2008) 028
  [arXiv:0806.0605 [hep-th]].
  
   
\bibitem{Chiodaroli:2009yw}
  M.~Chiodaroli, M.~Gutperle and D.~Krym,
  ``Half-BPS Solutions locally asymptotic to AdS(3) x S**3 and interface conformal field theories,''
  JHEP {\bf 1002} (2010) 066
  [arXiv:0910.0466 [hep-th]].
  
  
\bibitem{Chiodaroli:2011nr}
  M.~Chiodaroli, E.~D'Hoker, Y.~Guo and M.~Gutperle,
  ``Exact half-BPS string-junction solutions in six-dimensional supergravity,''
  JHEP {\bf 1112} (2011) 086
  [arXiv:1107.1722 [hep-th]].
  
  
\bibitem{Petkova:2000ip}
  V.~B.~Petkova and J.~B.~Zuber,
  ``Generalized twisted partition functions,''
  Phys.\ Lett.\ B {\bf 504} (2001) 157
  [hep-th/0011021].
  
\bibitem{Bachas:2001vj}
  C.~Bachas, J.~de Boer, R.~Dijkgraaf and H.~Ooguri,
  ``Permeable conformal walls and holography,''
  JHEP {\bf 0206} (2002) 027
  [hep-th/0111210].
  
\bibitem{Quella:2006de}
  T.~Quella, I.~Runkel and G.~M.~T.~Watts,
  ``Reflection and transmission for conformal defects,''
  JHEP {\bf 0704} (2007) 095
  [hep-th/0611296].
  
  
\bibitem{Gaiotto:2012np}
  D.~Gaiotto,
  ``Domain Walls for Two-Dimensional Renormalization Group Flows,''
  JHEP {\bf 1212} (2012) 103
  [arXiv:1201.0767 [hep-th]].

 
   
\bibitem{Gutperle:2012hy}
  M.~Gutperle and J.~Samani,
  ``Holographic RG-flows and Boundary CFTs,''
  Phys.\ Rev.\ D {\bf 86} (2012) 106007
  [arXiv:1207.7325 [hep-th]].
  
\bibitem{Oshikawa:1996dj}
  M.~Oshikawa and I.~Affleck,
  ``Boundary conformal field theory approach to the critical two-dimensional Ising model with a defect line,''
  Nucl.\ Phys.\ B {\bf 495} (1997) 533
  [cond-mat/9612187].
  
  
\bibitem{Calabrese:2004eu}
  P.~Calabrese and J.~L.~Cardy,
  ``Entanglement entropy and quantum field theory,''
  J.\ Stat.\ Mech.\  {\bf 0406} (2004) P06002
  [hep-th/0405152].
  
  
\bibitem{Affleck:1991tk}
  I.~Affleck and A.~W.~W.~Ludwig,
  ``Universal noninteger 'ground state degeneracy' in critical quantum systems,''
  Phys.\ Rev.\ Lett.\  {\bf 67} (1991) 161.
 
 
\bibitem{Azeyanagi:2007qj}
  T.~Azeyanagi, A.~Karch, T.~Takayanagi and E.~G.~Thompson,
  ``Holographic calculation of boundary entropy,''
  JHEP {\bf 0803} (2008) 054
  [arXiv:0712.1850 [hep-th]].
  
\bibitem{Chiodaroli:2010ur}
  M.~Chiodaroli, M.~Gutperle and L.~-Y.~Hung,
  ``Boundary entropy of supersymmetric Janus solutions,''
  JHEP {\bf 1009} (2010) 082
  [arXiv:1005.4433 [hep-th]].
  
\bibitem{Chiodaroli:2010mv}
  M.~Chiodaroli, M.~Gutperle, L.~-Y.~Hung and D.~Krym,
  ``String Junctions and Holographic Interfaces,''
  Phys.\ Rev.\ D {\bf 83} (2011) 026003
  [arXiv:1010.2758 [hep-th]].
  
  
\bibitem{Tan:2011tj}
  H.~-S.~Tan,
  ``Aspects of Three-dimensional Spin-4 Gravity,''
  JHEP {\bf 1202} (2012) 035
  [arXiv:1111.2834 [hep-th]].
 
  }
  
\end{thebibliography}
\end{document}